\documentclass[aps,amssymb,amsmath,twocolumn,prl,superscriptaddress]{revtex4}
\usepackage[pdftex]{graphicx}
\usepackage{amssymb, amsmath}
\usepackage{dcolumn}
\usepackage{bm}
\usepackage{bbm}
\usepackage{hyperref}
\usepackage{color}
\usepackage[up]{subfigure}
\usepackage{amsmath}
\usepackage{amsfonts}
\usepackage{amssymb}

\newcommand{\be}{\begin{equation}}
\newcommand{\ee}{\end{equation}}
\newcommand{\bc}{\begin{center}}
\newcommand{\ec}{\end{center}}
\newcommand{\bea}{\begin{eqnarray}}
\newcommand{\eea}{\end{eqnarray}}
\newcommand{\ba}{\begin{array}}
\newcommand{\ea}{\end{array}}

\newcommand{\singlespacing}{\let\CS=\@currsize\renewcommand{\baselinestretch}{1.5}\tiny\CS}
\newcommand{\doublespacing}{\let\CS=\@currsize\renewcommand{\baselinestretch}{1.35}\tiny\CS}

\def\@citex[#1]#2{\if@filesw\immediate\write\@auxout{\string\citation{#2}}\fi
  \def\@citea{}\@cite{\@for\@citeb:=#2\do
    {\@citea\def\@citea{,\linebreak[0]\hskip0pt plus .2em}%
      \@ifundefined{b@\@citeb}%
    {{\bf ?}\@warning{Citation `\@citeb' on page \thepage\space undefined}}%
      \hbox{\csname b@\@citeb\endcsname}}}{#1}}

\newtheorem{rule-def}[theorem]{Rule}

\begin{document}

\title{An Operational Meaning of Discord in terms of Teleportation Fidelity}
\author{Satyabrata Adhikari}
\email{satya@iitj.ac.in} \affiliation{Indian Institute of
Technology, Rajasthan, Jodhpur 342011, India}

\author{Subhashish \surname{Banerjee}}
\email{subhashish@iitj.ac.in}
\affiliation{Indian Institute of Technology, Rajasthan, Jodhpur 342011, India}

\begin{abstract}
Quantum discord is a prominent measure of quantum correlations, playing an important role
in expanding its horizon beyond entanglement. Here we provide an operational meaning of
(geometric) discord, which quantifies the amount of non-classical correlation of an
arbitrary quantum system in terms of its minimal distance from
the set of classical states, in terms of teleportation fidelity for general two qubit and $d \otimes d$ dimensional
isotropic and Werner states. 
A critical value of the discord is found beyond which the two qubit
state must violate the Bell inequality. This is illustrated by an open system model of a
dissipative two qubit. For the $d \otimes d$ dimensional states the
lower bound of  discord is shown to be obtainable from an experimentally measurable witness operator.
\end{abstract}


\maketitle

{\em Introduction.}-Quantum correlations occupy a central position in the quest
for understanding and harvesting the power of quantum mechanics. This point of view
has been highlighted in recent times by numerous developments in the
field of quantum information.  Entanglement \cite{wootters}, till about a decade back,
was considered synonymous with quantum correlations. This was a natural outcome of the
quest to understand the role of nonlocality in quantum mechanics, having a historical
lineage from Einstein-Podolsky-Rosen \cite{epr}, to Bell's inequality \cite{bell}, leading
to refinements resulting in the Bell-CHSH (Clauser-Horn-Shimony-Holt) inequalities \cite{chsh}.
With the advent of quantum discord \cite{ollivier, henderson}, the difference between the quantum
generalizations of two classically equivalent formulations of mutual information, realization
dawned that quantum correlations are bigger than entanglement. Thus, for example, separable
states, having by definition zero entanglement, could have non-zero discord. Also,
in the DQC1 model \cite{knill}, entanglement is negligible but there is sufficient amount of quantum discord
for a speed up over the best known classical algorithms.

Quantum discord for two qubit states with maximally mixed marginals was analytically obtained in
\cite{luo}, while in  \cite{girolami1} an algorithm  to calculate quantum discord for general two qubit states was developed.
In general, it is very difficult to obtain an analytical formula for quantum discord because it
involves an optimization over local measurements, requiring numerical methods. To
overcome this difficulty, another measure of quantum correlation called geometric discord was introduced in \cite{dakic}
which quantifies the amount of non-classical correlation of an
arbitrary quantum composite system in terms of its minimal distance from
the set of classical states. An analytical expression for geometric discord for the two qubit case was also found. This was generalized in \cite{lfu} to the
case of a $d \otimes d'$ dimensional system.
There is now an abundance of measures of quantum correlations such as quantum work deficit \cite{horo}, measurement induced disturbance \cite{luo1}
and dissonance \cite{modi}.

The understanding of a particular facet of a complex entity such as quantum correlations is greatly accentuated by the
development of various operational tasks to which it can be put to use. This has been the case, particularly, for entanglement
which was used for developing various useful aspects of quantum information processing such as teleportation \cite{bennett},
remote state preparation \cite{pati}, quantum cryptography \cite{ekert}, and quantum dense coding \cite{bennett1}.
Discord, likewise, has found use in explaining local broadcasting \cite{piani}, quantum state
merging \cite{madhok} and remote state preparation \cite{dakic1}. Here we provide an operational meaning of (geometric) discord
in terms of teleportation, the canonical model of quantum information and communication.  Teleportation is particularly
important, due not only to its operational aspect and experimental realization \cite{zeilinger} but also because of the fundamental role
it plays in sharpening understanding of Bell's inequality and entanglement \cite{gisin, rmhorodecki}.

In this letter, we establish a connection between geometric discord and teleportation fidelity
for general two qubit and $d \otimes d$ dimensional
isotropic  states. 
A critical value of the discord is found beyond which the two qubit
state must violate the Bell-CHSH inequality. This is illustrated by an open system model of a
dissipative two qubit \cite{sb10}. In addition, for $d \otimes d$ dimensional
isotropic  and Werner states, we develop lower bounds of geometric discord in terms of  experimentally measurable witness operators. 

{\em Maximum and Minimum value of Geometric Discord.}-
Any arbitrary two qubit mixed state can be written as
$\rho=\frac{1}{4}({{\mathbb I}}_{2}\otimes {{\mathbb I}}_{2} +
\vec{x}.\vec{\sigma}\otimes {{\mathbb I}}_{2} + {{\mathbb
I}}_{2}\otimes \vec{y}.\vec{\sigma} +
\sum_{i,j=1}^{3}t_{ij}\sigma_{i}\otimes\sigma_{j}).$
Here ${{\mathbb I}}_{2}$ is the two dimensional identity matrix,
$x_{i}=\mathrm{Tr}(\rho(\sigma_{i}\otimes {{\mathbb I}}_{2}))$,
$y_{i}=\mathrm{Tr}(\rho({{\mathbb I}}_{2}\otimes \sigma_{i}))$ are
components of local Bloch vectors $\vec{x}$ and $\vec{y}$,
respectively, while $\{t_{ij}\} \equiv T =  \mathrm{Tr}(\rho (\sigma_i
\otimes \sigma_j))$ denotes the correlation matrix and
$\sigma_{i}'s (i=1,2,3)$ are the Pauli matrices. The geometric
discord, normalized with respect to teleportation fidelity, is
defined as \cite{dakic}
$D_{G}(\rho)=\frac{4}{3} min_{\chi \in
\Omega_{0}}\|\rho-\chi\|_{2}^{2}$,
where $\Omega_{0}$ denotes the set of all zero discord states and
$\|.\|_{2}$ denotes the Hilbert-Schmidt norm and is defined as
$\|A\|_{2}=\sqrt{\mathrm{Tr}(AA^{\dagger})}$. For the case of two qubits, 
geometric discord was shown \cite{dakic} to be
$D_{G}(\rho)=
\frac{1}{3}[\|\vec{x}\|^{2}+\|T\|^{2}-\lambda_{max}(\vec{x}\vec{x}^{\dagger}+ T T^{\dagger})].$
Here $\lambda_{max}(\vec{x}\vec{x}^{\dagger}+ T T^{\dagger})$ is the maximum eigenvalue
of the matrix $\vec{x}\vec{x}^{\dagger}+ T T^{\dagger}$.
To proceed, we make use of a very useful theorem by Weyl \cite{horn}, which connects the
eigenvalues of the sum of Hermitian matrices to those of the individual matrices and is made use of in,
for example, understanding the stability of the spectrum of a Hermitian matrix with respect to perturbations. For convenience,
we present the theorem below.

\textbf{Theorem:} Let $X, Y \in M_{n}$ be Hermitian matrices and
let the eigenvalues $\lambda_{i}(X),\lambda_{i}(Y)$ and
$\lambda_{i}(X+Y)$ be arranged in an increasing order. For each
$k=1,2,...n$, we have
\begin{eqnarray}
\lambda_{k}(X)+\lambda_{1}(Y)\leq \lambda_{k}(X+Y)\leq
\lambda_{k}(X)+\lambda_{n}(Y), \label{inequality1}
\end{eqnarray}
where $M_{n}$ denotes the set of $n \times n$ Hermitian matrices
and $\lambda_{1}(Y),\lambda_{n}(Y)$ denotes the minimum and
maximum eigenvalues of Y, respectively. In particular, for $k=n$,
the inequality [Eq. (\ref{inequality1})] reduces to
$\lambda_{max}(X)+\lambda_{min}(Y)\leq\lambda_{max}(X+Y)\leq
\lambda_{max}(X)+ \lambda_{max}(Y).$
If we identify the Hermitian matrices $\vec{x}\vec{x}^{\dagger}$ and
$T T^{\dagger}$ with $X$ and $Y$, respectively, then this inequality gives
$\lambda_{max}(\vec{x}\vec{x}^{\dagger})+\lambda_{min}(TT^{\dagger})\leq\lambda_{max}(\vec{x}\vec{x}^{\dagger}+TT^{\dagger})\leq
\lambda_{max}(\vec{x}\vec{x}^{\dagger})+\lambda_{max}(TT^{\dagger}).$
Using this and the form of geometric discord for two qubits, we have
\begin{eqnarray}
\frac{1}{3}[\|\vec{x}\|^{2}+\|T\|^{2}-\lambda_{max}(\vec{x}\vec{x}^{\dagger})-\lambda_{max}(TT^{\dagger})]
\leq D_{G}(\rho) \nonumber\\\leq
\frac{1}{3}[\|\vec{x}\|^{2}+\|T\|^{2}-\lambda_{max}(\vec{x}\vec{x}^{\dagger})-\lambda_{min}(TT^{\dagger})].
\label{discordbound}
\end{eqnarray}
From these inequalities, the maximum and minimum value of $D_{G}(\rho)$ can be seen to be
\begin{eqnarray}
D_{G}^{min}(\rho)&=&\frac{1}{3}[\|T\|^{2}-\lambda_{max}(TT^{\dagger})],\\
D_{G}^{max}(\rho) &=& \frac{1}{3}[\|T\|^{2}-\lambda_{min}(TT^{\dagger})].
\label{maxandmindiscord1}
\end{eqnarray}
These results would be needed to connect quantum discord with Bell's inequality and teleportation.
If all the eigenvalues of the matrix $TT^{\dagger}$ are
equal then $\lambda_{max}(TT^{\dagger})=\lambda_{min}(TT^{\dagger})$ and we have
\begin{eqnarray}
D_{G}^{min}(\rho)=D_{G}^{max}(\rho)=D_{G}(\rho). \label{discord1}
\end{eqnarray}
An example where such an equality is realized is the Werner state \cite{werner}, a point to which we will return later.
The above obtained bounds can be used to calculate the maximum value of quantum discord
for separable states in $2 \otimes 2$ systems. Any separable state in a $2 \otimes 2$ system
can be expressed as
$\rho_{sep}= \sum_{k}p_{k}\frac{1}{4}(I\otimes
I+\vec{x}^{k}.\vec{\sigma}\otimes I+I\otimes
\vec{y}^{k}.\vec{\sigma}+
\sum_{i}x_{i}^{k}y_{i}^{k}\sigma_{i}\otimes\sigma_{i})$,
where, $x_{i}^{k},y_{i}^{k} \in \mathbb{R},
|\vec{x}^{k}|\leq1,|\vec{y}^{k}|\leq1$. 
Thus, for separable states, the correlation matrix $T$ is the
product of the two local Bloch vectors $\vec{x}$ and $\vec{y}$,
that is, $T=\vec{x}^{\dagger}\vec{y}$, where $\vec{x}=(x_{1},x_{2},x_{3})$
and $\vec{y}=(y_{1},y_{2},y_{3})$. The
maximum quantum discord $D_{G}^{max}$, for separable states, is
$D_{G}^{max}(\rho_{sep})= \frac{1}{3}(|x_{1}|^{2} + |x_{2}|^{2} +
|x_{3}|^{2})(|y_{1}|^{2} + |y_{2}|^{2} +
|y_{3}|^{2})$. 
Since $|\vec{x}^{k}|\leq1,|\vec{y}^{k}|\leq1$, so
$D_{G}^{max}(\rho_{sep})\leq \frac{1}{3}.$
It follows that for separable states $\rho_{sep}$, we have the inequality
\begin{eqnarray}
0 \leq D_{G}(\rho_{sep})\leq D_{G}^{max}(\rho_{sep}) \leq
\frac{1}{3}. \label{discordseparable}
\end{eqnarray}
Let us consider a state described by the density operator
$\rho_{1}= \frac{1}{4}[I\otimes I+\sigma_{x}\otimes I+I\otimes
\sigma_{x}+\sigma_{x}\otimes\sigma_{x}].$
It is separable since the partial transpose with respect to one of the qubits is a valid density matrix, that is,
$\rho_{1}^{T_{A}}=\rho_{1}$. In this case, the maximum value of discord is
given by $D_{G}^{max}(\rho_{1})= \frac{1}{3}.$
Thus we show that there exist separable states for which the upper
bound of the maximum discord is achieved.

{\em Relation between Quantum Discord, Bell's inequality and Teleportation Fidelity.}-
We now establish a relation between
maximum discord and teleportation fidelity, which also gives an
operational meaning of quantum discord. To achieve our goal, let
us consider
\begin{eqnarray}
D_{G}^{max}(\rho)=
\frac{1}{3}[\|T\|^{2}-\lambda_{min}(TT^{\dagger})] =
\frac{1}{3}M(\rho),\label{maxdiscord}
\end{eqnarray}
where $M(\rho)=max_{i>j}(u_{i}+u_{j})$, $u_{i},u_{j}$ are the
eigenvalues of $TT^{\dagger}$.\\
If the state described by the density matrix $\rho$ satisfies the Bell-CHSH inequality then it follows that
\begin{eqnarray}
D_{G}^{max}(\rho)\leq \frac{1}{3}. \label{maxdiscordbellineqality}
\end{eqnarray}
The above result holds for all separable states as is evident from Eq. (\ref{discordseparable}).
But there may also exist some entangled states that satisfy it.

\textbf{Theorem-1:} Any two qubit mixed state $\rho$ is entangled
if $D_{G}^{max}(\rho)>\frac{1}{3}$.

\textbf{Proof:} Any two qubit mixed state $\rho$ is
entangled if it violates the Bell inequality. The Bell inequality
is violated iff $M(\rho)>1$ \cite{rhorodecki}. Therefore
the theorem follows from Eq. (\ref{maxdiscord}).

To establish an operational meaning of discord, we use a series of connections, that is,
the relation of discord with negativity and then negativity with teleportation fidelity.
The relation between an entanglement measure known as negativity
($N(\rho^{ent})$) and quantum discord $D_{G}(\rho^{ent})$ for an
entangled state $\rho^{ent}$ is given by \cite{girolami}
$ N^{2}(\rho^{ent})\leq D_{G}(\rho^{ent})$.
Let $S=\{\rho^{ent}_{CHSH}:D_{G}^{max}(\rho^{ent}_{CHSH})\leq
\frac{1}{3}\} $ denotes the set of entangled states which satisfy
the Bell-CHSH inequality. Therefore the states which belong to the
set $S$ must satisfy the inequality
$ N^{2}(\rho^{ent}_{CHSH})\leq D_{G}(\rho^{ent}_{CHSH})\leq
D_{G}^{max}(\rho^{ent}_{CHSH})\leq \frac{1}{3}.$
For any two qubit mixed state $\rho$, teleportation fidelity
$F(\rho)$ is related to negativity as \cite{verstraete}
$ 3F(\rho)-2\leq N(\rho)$. 
Since this inequality holds for all
$\rho^{ent}_{CHSH}$, we have
$ (3F(\rho^{ent}_{CHSH})-2)^{2}\leq
D_{G}^{max}(\rho^{ent}_{CHSH})\leq \frac{1}{3}. $
From Eq. (\ref{maxdiscordbellineqality}) and the relations, discussed above, 
connecting negativity with discord and teleportation fidelity,
it follows that $\frac{2}{3}<F(\rho^{ent}_{CHSH}) \leq \frac{2}{3}+\frac{1}{3\sqrt{3}}$.  This is what is
expected for states satisfying Bell's inequality, but at the same time useful for teleportation, bringing
out the consistency of our results.
Thus we obtain a bound of
quantum correlation measured by discord for those entangled
states which satisfy Bell's inequality but are still useful
for teleportation. This result can be expressed in the form of the following theorem.

\textbf{Theorem-2:} If the entangled state $\rho$ satisfies Bell's inequality but is useful for teleportation then $D_{G}^{max}(\rho)$
must satisfy the inequality
\begin{eqnarray}
(3F(\rho)-2)^{2}\leq D_{G}^{max}(\rho)\leq \frac{1}{3}.
\label{negativitydiscordbelltel1}
\end{eqnarray}
As $u_{i}\leq1$ $for~~ i=1,2,3$ and
$U(\rho)=\sum_{i=1}^{3}\sqrt{u_{i}}$, $M(\rho)$ and $U(\rho)$ are related as
$M(\rho)\leq U(\rho)$ \cite{rmhorodecki}, while the relation between $U(\rho)$ and teleportation fidelity
$F(\rho)$ is $F(\rho)=\frac{1}{2}[1+\frac{1}{3}U(\rho)]$ \cite{rmhorodecki}. Using these relations, Eq. (\ref{maxdiscord}) can be seen to reduce to
$ D_{G}^{max}(\rho)\leq 2F(\rho)-1$.
Since this inequality  holds for any mixed two qubit
state, states not useful for teleportation must satisfy
\begin{eqnarray}
0\leq D_{G}^{max}(\rho)\leq 2F(\rho)-1,~~F(\rho)\leq\frac{2}{3}.
\label{maxdiscordnotuseful}
\end{eqnarray}
\textbf{Theorem-3:} A two qubit state $\rho$ violates Bell-CHSH inequality and is useful for
teleportation iff
\begin{eqnarray}
\frac{1}{3}<D_{G}^{max}(\rho)\leq 2F(\rho)-1,~~F(\rho)>\frac{2}{3}.
\label{maxdiscord2}
\end{eqnarray}
\textbf{Proof:} It follows from \textbf{Theorem-1} and the inequality, shown above, connecting the maximum value of discord
to the teleportation fidelity.

Eqs. (\ref{negativitydiscordbelltel1}), (\ref{maxdiscordnotuseful})
and (\ref{maxdiscord2}) are our principal results for the two qubit states, providing the relationship
between discord and teleportation fidelity and covers the regimes where Bell's inequality is satisfied but
the states may or may not be useful for teleportation as well as where the inequality gets violated, the critical value of
discord above which Bell's inequality gets violated being $1/3$.

\textbf{Illustrations:}
(a). Let us consider a Werner state $\rho_{W} = p|\psi^{-}\rangle\langle\psi^{-}|+(1-p)\frac{I}{4}$,~~$0\leq
p\leq1$, where $|\psi^{-}\rangle=\frac{1}{\sqrt{2}}(|01\rangle-|10\rangle)$ is the singlet state.
The correlation matrix $T$ for $\rho_{W}$ is given by
$T=\left(%
\begin{array}{cccccccc}
  -p & 0 & 0 \\
  0 & -p & 0 \\
  0 & 0 & -p \\
  \end{array}%
\right).$
As all the eigenvalues of $TT^{\dagger}$ are  equal, we have
$D_{G}^{min}(\rho_{W})=D_{G}^{max}(\rho_{W})=D_{G}(\rho_{W})$, thus realizing Eq. (\ref{discord1}).
The quantum discord is
$D_{G}(\rho_{W})=\frac{2}{3}p^2.$ 
From Eq. (\ref{maxdiscordbellineqality}), it is clear that $\rho_{W}$
violates Bell inequality iff $p>\frac{1}{\sqrt{2}}$. Also, using $F(\rho_{W}) = (1+p)/2$ \cite{tapi}
in Eq. (\ref{negativitydiscordbelltel1}), it is easy to see that Werner states satisfying Bell's inequality, but still
useful for teleportation have the parameter $p$ in the range $\frac{1}{3} < p \le \frac{1}{3} + \frac{2}{3\sqrt{3}}$.
This is known in the literature, providing a nice consistency check of our results. Finally, it follows from Eq. (\ref{maxdiscord2}) that
$\rho_{W}$ is always useful for teleportation iff
$\frac{1}{3}<D_{G}(\rho_{W})\leq p,~~~p>\frac{1}{3}.$

\begin{figure}[ht]
\includegraphics[width=6.0cm]{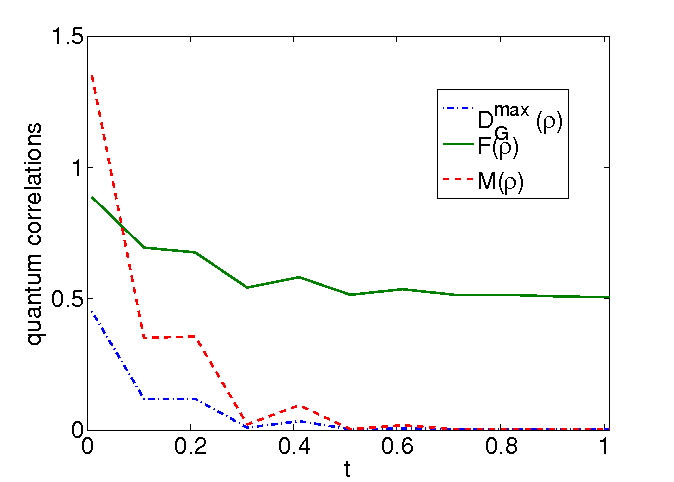}
\caption{(Color online) The figure depicts quantum correlations; Bell's inequality $M(\rho)$, teleportation fidelity $F(\rho)$ and maximum value of 
geometric discord $D_{G}^{max} (\rho)$,
with respect to the time of evolution $t$. 
Here temperature (in units where $\hbar \equiv k_B = 1$) $T = 10$, inter qubit distance $r_{12} = 0.01$ (collective decoherence) and bath squeezing parameter $r = -1$. 
With time, states violating
Bell's inequality start satisfying it, a natural consequence of the degradation of quantum correlations due to open system effects.  At $t = 0.12$,  Bell's
inequality is violated, it can be seen that  Eq. (\ref{maxdiscord2}) gets satisfied. Again at $t = 0.2$, we have the case where Bell's inequality is satisfied but the
states are useful for teleportation, as seen by the inequality 
Eq. (\ref{negativitydiscordbelltel1}) being satisfied. Finally, for longer evolutions, such as $t = 0.6$, the states are no longer useful for teleportation, and the
figure satisfies Eq. (\ref{maxdiscordnotuseful}).}
\label{fig:1}
\end{figure}

(b). Next we consider an open quantum system model for two qubit mixed states \cite{sb10}, which could be thought of as a quantum
channel used for studying quantum correlations. Open quantum system is the study
of the evolution of a system of interest, such as a qubit, taking into account the effect of its surroundings alternatively called
reservoir or bath.  The evolution is mixed, in general, with decoherence and dissipation appearing as natural outcomes. We consider
two qubits interacting with a bath, modeled as an electromagnetic field in a squeezed thermal state, via the dipole interaction. The
system-reservoir coupling constant is dependent upon the position of the qubit, leading to interesting dynamical consequences. Basically
this allows a classification of the dynamics into two regimes: the independent decoherence regime, where the inter qubit distances are such that
each qubit sees an individual bath or the collective decoherence regime, where the qubits are close enough to justify a collective interaction
with the bath. In Fig.\,\ref{fig:1}, the evolution of Bell's inequality, teleportation fidelity and maximum value of geometric discord with respect to 
time is shown. It is very satisfactory that a dynamical model, which could be envisaged in an experimental setup \cite{ficek}, satisfies all the
inequalities developed above connecting discord to teleportation fidelity, that is, Eqs. (\ref{negativitydiscordbelltel1}), (\ref{maxdiscordnotuseful})
and (\ref{maxdiscord2}), as well as the critical value of discord ($=1/3$) above which the Bell-CHSH inequality gets violated.

{\em Relation between discord, teleportation fidelity and witness
operator in $d \otimes d$ system.}~ Here, we study  $d \otimes d$
systems and generalize the relation between quantum discord and
teleportation fidelity for states with symmetry, in particular,
the isotropic and Werner states and also derive experimentally
achievable bounds of discord for them, using witness operators. A
witness operator is a Hermitian operator which can detect
entangled states and  can be realized experimentally. Recently it
has been shown that the witness operator also takes part in discriminating states useful for teleportation \cite{nirman}. \\

The isotropic state is defined by
$ \rho_{f}=\frac{1-f}{d^2-1}(I-|\phi^{+}\rangle\langle\phi^{+}|)+f|\phi^{+}\rangle\langle\phi^{+}|$,
where
$|\phi^{+}\rangle=\frac{1}{\sqrt{d}}\sum_{i=0}^{d-1}|ii\rangle$, $f$ is the singlet fraction and $d$ is the dimension of an
individual component of the bipartite system.
Its negativity is \cite{lee}
$ N(\rho_{f}) =\frac{fd-1}{d-1}$ and zero, for $f>\frac{1}{d}$, or $f<\frac{1}{d}$, respectively. 
For an isotropic state $\rho_{f}$, the following statements are equivalent,
(a). $\rho_{f}$ is separable iff $\rho_{f}$ is PPT, (b). $\rho_{f}$ is PPT iff $0\leq f\leq \frac{1}{d}$.
Using the relation between negativity and discord as well as the connection 
of teleportation fidelity with singlet fraction
$F(\rho_{f})=\frac{df(\rho_{f})+1}{d+1}$ \cite{horo99}, we have
\begin{eqnarray}
(\frac{(d+1)F(\rho_{f})-2}{d-1})^{2}\leq
D_{G}(\rho_{f}),~~~F(\rho_{f})>\frac{2}{d+1}.
\label{teldiscordisotropicstate}
\end{eqnarray}
Eq. (\ref{teldiscordisotropicstate}) provides a theoretical
lower bound of discord as a function of the teleportation
fidelity. A question then naturally arises, can we also have a
lower bound which could be achieved in an experiment? The answer
is in the affirmative. The lower bound of quantum discord for
isotropic states can be achieved by witness operators. Witness
operators act as a hyperplane separating  entangled and
separable states and can be divided into two different classes:
decomposable and non-decomposable witness operators. Although
non-decomposable witness operators detect both negative partial
transpose (NPT) and positive partial transpose (PPT) entangled
states, decomposable witness operators can only detect NPT
entangled states. Recently, a generalized form of
optimal teleportation witness operator was proposed and showed to be a
decomposable entanglement witness operator \cite{atul}.

Let us consider a decomposable witness operator of the form
$W_{f}=\frac{1}{d}I-|\phi^{+}\rangle\langle\phi^{+}|$. Its expectation
value in the state $\rho_{f}$ is
\begin{eqnarray}
\mathrm{Tr}(W_{f}\rho_{f})=\frac{1}{d}-f,~~~~f>\frac{1}{d}.
\label{witnessisotropicstate}
\end{eqnarray}
Thus, every entangled isotropic state is detected by $W_{f}$.
Making use of Eq. (\ref{witnessisotropicstate}) and the negativity of an isotropic state, the lower bound of discord,
as a function of the witness operator $W_{f}$,  is
\begin{eqnarray}
\frac{d^{2}}{(d-1)^{2}}(-\mathrm{Tr}(W_{f}\rho_{f}))^{2}\leq
D_{G}(\rho_{f}). \label{lbwitnessisotropicstate}
\end{eqnarray}
Let us now consider the Werner state in a $d\otimes d$ dimensional
Hilbert space where it is defined as \cite{lee}
$ \rho_{x}=\frac{2(1-x)}{d(d+1)}(\sum_{k=0}^{d-1}|kk\rangle\langle
kk|+\sum_{i<j}|\Psi_{ij}^{+}\rangle\langle\Psi_{ij}^{+}|)+
\frac{2x}{d(d-1)}\sum_{i<j}|\Psi_{ij}^{-}\rangle\langle\Psi_{ij}^{-}|$.
Here $|\Psi_{ij}^{\pm}\rangle=\frac{1}{\sqrt{2}}(|ij\rangle\pm
|ji\rangle)$ and
$x= \mathrm{tr}(\rho_{x}\sum_{i<j}|\Psi_{ij}^{-}\rangle\langle\Psi_{ij}^{-}|)$.
Analogous to an isotropic state, the following statements are
equivalent for Werner state, (a). $\rho_{x}$ is separable iff
$\rho_{x}$ is PPT, (b). $\rho_{x}$ is PPT iff $0\leq x \leq
\frac{1}{2}$. The negativity of Werner state is
$N(\rho_{x})=\frac{2}{d}\frac{2x-1}{d-1}$ \cite{lee}, while it is
related to the singlet fraction as
$f(\rho_{x})\leq\frac{1+2N(\rho_{x})}{d}$ \cite{vidal}. If the
state $\rho_{x}$ is entangled, that is, $N(\rho_{x})\neq 0$, then
it is clear that every entangled state is useful for
teleportation. In this case, the lower bound of discord for the
Werner state, in terms of the witness operator $W_{x} =
(|\Psi\rangle\langle\Psi|)^{T_{A}},
\textrm{where}~~|\Psi\rangle=\frac{1}{\sqrt{d}}\sum_{i=0}^{d-1}|ii\rangle
$, is given by
\begin{eqnarray}
\frac{4}{(d-1)^{2}}(-\mathrm{Tr}(W_{x}\rho_{x}))^{2}\leq D_{G}(\rho_{x}).
\label{lbwitnesswernerstate}
\end{eqnarray}

Thus for the cases of the isotropic as well as the Werner states, a lower bound of discord is obtained by two different routes.
The first one, in theme with our approach for the $2 \otimes 2$ dimensional systems, goes about establishing a relation between
discord and teleportation fidelity by connecting the concepts of negativity, singlet fraction, teleportation fidelity and discord;
while the second approach relies upon the construction of decomposable witness operators.

For an experimental realization of the witness operation it is necessary to
decompose the witness into operators that can be measured locally,
that is, a decomposition into projectors of the form $W =
\sum_{i=1}^{k}c_{i}|e_{i}\rangle\langle e_{i}|\otimes
|f_{i}\rangle\langle f_{i}|$. The decomposition of the witness
operator $W_{f}$ for two qubit systems has been shown in
\cite{nirman}. Moreover, the witness operator $W_{x}$ can easily
be decomposed into local Pauli operators $\sigma_{i}, i=1,2,3$ and
local Gell-Mann matrices for two qubits and two qutrits systems,
respectively. It can be also extended to generalized Gell-Mann
matrices, which are standard $SU(d)$ generators, for d-dimensional
systems.

{\em Conclusions.}~In this work we have provided an operational
meaning of discord by connecting it to teleportation fidelity. In
$2 \otimes 2$ systems, we make use of a theorem of Weyl for
Hermitian matrices which proves to be the key threading together
the various links to provide the connection between discord
and teleportation fidelity. The results are seen to be consistent
when applied to the Werner state and a dynamically generated two
qubit open system model. This study is further extended to a higher dimensional $d
\otimes d$ isotropic system. We also obtain lower bounds of discord, 
for $d \otimes d$ isotropic and Wener states,
in terms of appropriate witness operators, and discuss how these can
be achieved experimentally. We hope this work motivates further research
into providing an operational meaning of discord in general higher dimensional systems
thereby harnessing its potential use in quantum information processing.

\end{document}